\documentclass[a4paper,16pt,epsf,amssymb,citesort]{article}

\usepackage{tabularx}
\usepackage{array}
\usepackage{graphics}
\usepackage{graphicx}
\usepackage{epsfig}
\usepackage{amsmath}
\usepackage{amssymb}
\usepackage{axodraw}
\usepackage{cite}
\usepackage{multirow}
\usepackage{verbatim}
\newcommand{\pslash}{p \!\!\!/}

\begin{document}
\date{}
\def\rR{{\rm R}}
\title{\bf Quark pseudoscalar vertex and quark mass function with clover fermions~: spontaneous symmetry breaking, OPE, symmetry restoration at small volume }
\author{ Ph.~Boucaud$^a$,
J.-P.~Leroy$^a$, A.~Le~Yaouanc$^a$, J. Micheli$^a$,\\ O.~P\`ene$^a$, J.~Rodr\'{\i}guez--Quintero$^b$}

\maketitle
\begin{flushright}
\begin{tabular}{l}
{\tt LPT Orsay 09-89}\\
{\tt UHU-FT/09-023}\\
\end{tabular}
\end{flushright}
\begin{center}
$^a$Laboratoire de Physique Th\'eorique\footnote{Unit\'e Mixte de Recherche 8627 du Centre National de la Recherche Scientifique},\\
Universit\'e de Paris XI, B\^atiment 210, 91405 Orsay Cedex,
France \\
$^b$ Dpto. F\'isica Aplicada, Fac. Ciencias Experimentales,\\
Universidad de Huelva, 21071 Huelva, Spain.
\end{center}


\begin{abstract}
We study the quark mass function on hypercubic lattices, in a large range of physical volumes and cutoffs. To avoid  the very large Wilson term artefact, we  exploit the relation between the quark mass function and  the pseudoscalar vertex  in the continuum.  We extrapolate to the chiral limit. 

In function of the physical volume, we observe a striking discontinuity in the properties of chiral extrapolation around a physical volume $L_c \simeq 6~{\mathrm GeV}^{-1}=1.2~fm$. It is present in the quark mass function, which collapses to zero, as well as in the  pion mass and the quark condensate as directly calculated from the pseudoscalar correlator. It is strongly reminiscent of the phenomenon of chiral symmetry restoration observed by Neuberger and Narayanan at $N_C=\infty$ around the same physical length.

In the case of spontaneous symmetry breaking, we confirm that the OPE of the quark mass function, involving the quark condensate,  is not operative at the available momenta, even taking into account the unusually large high order corrections 
to the Wilson coefficient calculated by Chetyrkin and Maier ; the gap  remains large, around a factor 2, even at the largest momenta available to us ($p\simeq 6~{\mathrm GeV}$). 

\end{abstract}

\section{Introduction}
\hspace{\parindent} Obviously, when chiral symmetry is spontaneously broken, the quark pseudoscalar vertex presents a Goldstone pole, i.e. a pole  at $q^2=m_{\pi}^2$. In the Euclidean region tested by lattice QCD, the large magnitude of its contribution to the pseudoscalar vertex at $q=0$ was noted first in numerical calculation by ref. \cite{martinelli9411010}. However, the denominator is $q^2+m_{\pi}^2$~; therefore it vanishes at $q=0$ in the chiral limit and it is proportional to the current quark mass~; we have then by exception a divergence of the vertex in the region tested by lattice QCD, i.e. it goes to infinity when $m_q \to 0$,
for all momenta $p$ of the quark legs ($q=p'-p=0$), a phenomenon which would be very spectacular if we could indeed approach very small quark masses on the lattice. One must note that in spite of this divergence, a chiral limit of the vertex was presented in the literature for quite a while in the context of MOM non perturbative renormalisation at zero mass \footnote{Let us stress that no such Goldstone contribution is expected in the method of renormalisation of the ALPHA group at zero mass.}. Attention to this divergence was drawn by the JLQCD collaboration \cite{aoki9901019}, and around the same time in \cite{cudell9810058,cudellNPPS83}. In the paper \cite{cudell9810058}, the problem was shown to have two distinct aspects~:

1) Usually, the pseudoscalar vertex is used to renormalise the pseudoscalar density in the UV asymptotic region~; one uses first  a numerical MOM renormalisation constant $Z_P^{MOM}$ to be later inserted into certain perturbative calculations (running to higher scales, scheme conversions). It is then necessary to {\bf subtract} the Goldstone boson pole, in order to extract the purely perturbative part of the renormalisation constant. Analogously, one should extract also any power corrections originating in condensates, which will be the case for the quark condensate contribution, when determining $Z_m$, see below, or for the $A^2$ condensate contribution in the gluon and ghost Green functions when extracting perturbative informations like $\Lambda_{QCD}$, see for instance \cite{boucaud08112059} and references thereins. 

2) However, as a second aspect,  one may also consider the Goldstone pole contribution not as a parasitic contribution to be eliminated, but in its own physical interest and in its relation with other interesting quantities. We recall these relations, stressed in the same paper. First, the residue of the pole is simply given, up to a finite renormalisation, by the {\bf quark mass function}, i.e. the quotient of the scalar part of the quark propagator and the vector part, in the chiral limit. That this mass function is non zero in the chiral limit is therefore also a signal of spontaneous symmetry breaking. This can be seen directly from the fact that the scalar part of the propagator is zero in a chirally symmetric vacuum. Then, this mass function can be said to constitute a {\bf continuous set of renormalised order parameters}, labelled by $p$. 

Finally, the spontaneous breaking also manifests itself through the OPE of this quark mass function~; the main contribution in the chiral limit is the one of the {\bf quark condensate}, with power $1/p^2$~; this gives therefore a relation with another, well known, indicator of spontaneous chiral symmetry breaking.

The object of the present paper is to study in more detail these relations on the lattice, at $N_f=0$, with the clover action. The central object will be the quark mass function, in the chiral limit, calculated, along the line described above, through the residue of the Goldstone boson in the pseudoscalar vertex.
Thus the present paper completes the study of the papers \cite{boucaud0307026,boucaud0504017} devoted to the vector part of the propagator, $Z_{\psi}$. For an extensive study of the quark propagator, one can refer also to the papers of the Adelaide group, see for instance our reference below \cite{zhang0301018}.

 Although it may appear involved at first, this method of calculation is indeed an advantageous approach in calculations with the clover action, because the lattice quark propagator itself would yield an untractable mass function, plagued with the huge Wilson term artefact, while the pseudoscalar vertex has much less artefacts, as will be seen. 

On the other hand, this calculation is very simple and fast even on large volumes in contrast to the calculations with overlap action. Of course, one cannot claim to work very close to the chiral limit. It is however instructive to see what comes out when using the standard methods of chiral extrapolation, successfully used since a long time
in phenomenological applications of lattice QCD.

With respect to the exploratory study presented in Ref. (\cite{cudell9810058,cudellNPPS83}) with the help of data from the QCDSF group, the present study uses the same basic ideas, but it is performed with better data, with various $\beta$'s and lattice sizes, so that for instance artefacts can be identified and OPE can be tested with larger momenta.

\subsection{Physics results}

\hspace{\parindent}{\bf OPE}~: let us recall, with the lattice data used in \cite{cudell9810058}, there appears a very large discrepancy between the prediction of OPE and the lattice data, using the known quark condensate value and the Wilson coefficient with the two calculated orders~; the prediction of the OPE  is found to be much lower than the lattice data. This problem with OPE is confirmed (see section \ref{OPEprincipal}). It can be partly elucidated thanks to the new perturbative calculations of Chetyrkin and Maier for the Wilson coefficient \cite{chetyrkin09110594,maier}~: the high order corrections to the Wilson coefficient are exceedingly large, almost spoiling the hope to extract the condensate from the high momentum lattice data. The discrepancy is admittedly reduced, but remains around a factor $2$ at the largest available momenta. 

{\bf Chiral symmetry ``restoration"}~: one also displays (see section \ref{discontinuity}) a new striking and unexpected phenomenon at small volumes, that we can term for simplicity as an abrupt  ``chiral symmetry restoration" below some critical {\bf physical} volume~; it affects the chiral extrapolation of several quantities at the same time, not only the quark mass function, but also the pion mass and the quark condensate~; to evaluate the latter independently of the GMOR (Gell-Mann, Oakes, Renner) identity, we recourse to the old method proposed in the paper by Bochicchio et al. \cite{bochicchio} . This discontinuity is reminiscent of the phase transition found by Narayanan and Neuberger \cite{narayanan} at $N_c \to \infty$. Let us recall at this point that usually it is expected that the chiral symmetry order parameters are vanishing at zero quark mass, in finite volumes~; therefore, what one should observe on the lattice is depending on the physical volume and on the smallest quark masses actually considered.

\section{Setting the  framework in the continuum~: the pseudoscalar vertex, the W-T identity and the quark mass function} \label{continuum}

\subsection{Definitions and Lorentz invariance}
\hspace{\parindent}We work in the Landau gauge.
Let us first fix the notations that we will use.
We will use all along the Euclidean metrics.  The continuum quark propagator
is a $12 \times 12$ matrix $S(p)$ for 3-color and 4-spinor indices. One can take into account Lorentz (in fact $O(4)$) invariance and discrete symmetries, as well as color neutrality of the vacuum by expanding the inverse propagator according to~:
\begin{eqnarray}\label{Sm1}
S^{-1}(p) = \delta_{a,b} Z_\psi(p^2) \left( i\,\pslash + m(p^2)\right)
\end{eqnarray}
where $a,b$ are the color indices. $Z_\psi(p^2)$ is a standard lattice notation , referring to the role it plays as a renormalisation constant for the quark field (in the standard Georgi-Politzer MOM renormalisation, see subsection \ref{MOM},  $Z_2^{MOM}(\mu^2)=Z_\psi(\mu^2)$~; for the precise lattice definition, see below, section \ref{lattice}).
Obviously, one has in the continuum, with trace on spin and color~:
\begin{eqnarray}\label{zpsi}
Z_\psi(p^2)=1/i~1/12~Tr(S^{-1}(p)\pslash)/p^2 
\end{eqnarray}
Sometimes, one uses to describe the scalar part of the propagator, instead of the quark mass function $m(p^2)$, the alternative quantity, which is the proper scalar part of the bare propagator~:
\begin{eqnarray}\label{b}
b(p^2) = Z_\psi(p^2) \, m(p^2).
\end{eqnarray}
One must not forget however the advantage of $m(p^2)$ being {\bf UV finite}, since it is the ratio of two quantities renormalised by the same factor $Z_2$. In fact it is the MOM renormalised quark mass, with renormalisation point $\mu^2=p^2$ (see below). 

This is the central physical quantity we will consider in the present paper. On the other hand,we will
consider as auxiliary the pseudoscalar vertex, and then use the axial Ward identity to relate this vertex to $m(p^2)$. Indeed, it happens that $m(p^2)$ is not directly calculable with the clover action. We then need some more
definitions.

Let us consider a colorless local two quark operator $\bar q {\cal O} q$.
 The corresponding three point Green function  $G$ is defined by
\begin{eqnarray}\label{G}
G(p, q) = \int d^4x d^4y~e^{i p \cdot y + i q \cdot x} 
< q(y)\bar q(x) {\cal O} q(x) \bar q(0) >
\end{eqnarray}
It is a $4 \times 4$ matrix in Dirac space.
The corresponding vertex function is then defined by amputation
of quark propagators on both sides~:
\begin{eqnarray}\label{gammamu1}
\Gamma(p, q) = S^{-1}(p) \,G(p, q)\,S^{-1}(p+q)
\end{eqnarray}
In the whole  paper, we will restrict ourselves to the case where
the operator carries a vanishing momentum transfer $q_\mu=0$. In the following we will omit to write $q_\mu=0$ and we will moreover understand
$\Gamma(p)$ as the {\bf bare} vertex function computed on the lattice.

Now, Lorentz covariance and discrete symmetries allow to write for the axial vertex~:
\begin{eqnarray}\label{gammamu2}
\Gamma_{A \mu}(p) = \delta_{a,b} [g_A^{(1)}(p^2) \gamma_\mu  \gamma_5+
 i g_A^{(2)}(p^2) p_\mu  \gamma_5 + \nonumber \\
g_A^{(3)}(p^2) p_\mu \pslash  \gamma_5+ ig_A^{(4)}(p^2) [\gamma_\mu,\pslash] \gamma_5 ]
\end{eqnarray}
which should be obeyed approximately on the lattice, as we checked, and similarly for the pseudoscalar vertex~:.
\begin{eqnarray}\label{gamma5}
\Gamma_(p^2) = \delta_{a,b} \left[ g_5^{(1)}(p^2) \gamma_5 + i g_5^{(2)}(p^2) \gamma_5 \pslash \right]
\end{eqnarray}

\subsection{Renormalisation and Ward-Takahashi Identities}
\hspace{\parindent}Although we do not require a specific renormalisation scheme, we have to discuss the renormalisation, because the Ward-Takahashi(W-T) identities should be imposed on the renormalised theory, and not on the bare quantities (we do not consider anomalies). The following considerations hold in an arbitrary renormalisation scheme, and the corresponding renormalised quantities are denoted by a sub- or superindex  ${\rm R}$. The renormalised equations should hold up to ${\cal O}(a)$ artefacts. 

$Z_2$ denotes as usual the fermion field or propagator renormalisation according to~:
\begin{eqnarray}
q=\sqrt{Z_2} q_{\rm R} \nonumber \\
S(p)= Z_2 S_{\rm R}(p)
\end{eqnarray}
Let us recall that the corresponding renormalised vertex functions are ~:
\begin{eqnarray}
\Gamma(p)=Z_2^{-1} Z_{\cal O}^{-1} \Gamma_{\rm R}(p) , 
\end{eqnarray}
where the necessary subindices are implicit for each type of vertex~; $Z_{\cal O}$ is the renormalisation of the composite operator, namely a current or density operator~: ${\cal O}=j_V,j_A,P_5$~; the $Z_2$ factor takes  into account the amputation . 

Note that the standard definition of renormalisation constants is to \underline{divide} the bare quantity by the renormalisation constant to obtain the renormalised quantity (except for photon or gluon vertex renormalisation factors $Z_1$ which we do not use). In principle, renormalisation of composite operators, for instance $Z_V$, should be defined similarly. We have followed this convention in our works on gluon fields, for the renormalisation of the gauge dependent gluon field condensate $A^2$. But, in the case of \underline{quark} composite operators,
an opposite convention has become standard in lattice calculations~: $(\bar q {\cal O} q)_{bare}=Z_{\cal O}^{-1} (\bar q {\cal O} q)_{\rm R}$~; we feel compelled to maintain this convention for the sake of comparison with parallel works on the lattice. This explains our writing of the renormalised vertex function. 

In the continuum limit $Z_V=1$, and in the chiral limit, $Z_A=1$ (conserved currents). We keep $Z_A$ since the axial current is not conserved away from the chiral limit, which we take only in the end.  The lattice artefacts, as any  other regularisation scheme, generate finite ${\cal O}(g^2)$ effects, vanishing slowly with $a$, included in factors $Z_V,~Z_A$, due to additional divergencies multiplying the $a$ terms (which have higher dimension). There are also terms with powers of $a$ which we do not write. $Z_V,~Z_A$ are independent of the renormalisation scheme up to such terms. The fact that we do not include such terms means that our equations should hold only sufficient close to the continuum.

{\bf Consequences of the axial Ward identity} Let us develop the consequences of the axial W-T identity, which derives from the equation $\partial_{\mu} (j_A)^{\mu}= 2m P_5$.  For this purpose, one  has to return momently to the general case $q=p'-p \neq 0$.
Since they reflect the symmetries of the physical theory, the naive Ward identities should a priori hold for the renormalised Green functions (except for anomalies) and at infinite cutoff, which means~: 
\begin{eqnarray}\label{AwardR}
q_{\mu}\ \Gamma_{A\mu,{\rm R}}(p,q)=-i\,(S_{\rm R}^{-1}(p+q)\gamma_5 +\gamma_5 S_{\rm R}^{-1}(p))+i \,2 m_{\rm R} \Gamma_{5,{\rm R}}(p,q)
\label{AWTIR}
\end{eqnarray}
$m_{\rm R}$ is the renormalised mass in the scheme, $m_{\rm R}=Z_m^{-1} m_q$. It is then possible to return to bare quantities which are the ones actually measured on the lattice. 
 Multiplying both sides by  $Z_2^{-1}$ one gets~:
\begin{eqnarray}\label{AwardB}
Z_A~q^{\mu}\Gamma_{A\mu}(p,q) = -i\, (S^{-1}(p+q)\gamma_5+\gamma_5 S^{-1}(p))+i\,2 m_{\rm R} Z_P \Gamma_5(p,q),\quad.
\end{eqnarray}
i.e. there appear the renormalisation constants  $Z_A,Z_P$.
Now, we exploit this equation through an expansion in powers of $q^{\mu}$.

{\bf Setting $q=0$. The pseudoscalar vertex.} First, from (\ref{Sm1})-(\ref{gammamu2}), one finds constraints for the pseudoscalar vertex by making $q_{\mu}=0$, which eliminates the axial vertex~:
\begin{eqnarray}\label{Award3}
m(p^2)Z_{\psi}(p^2)\gamma_5=m_{\rm R} Z_P \Gamma_5(p)  
\end{eqnarray}
This means that the pseudoscalar vertex contains only the $\gamma_5$ component, i.e. the second term in (\ref{gamma5}) should vanish~:
\begin{eqnarray}
g_5^{(2)}(p^2)=0 \nonumber \\ 
\Gamma_5(p) = g_5^{(1)}(p^2) \gamma_5.
\end{eqnarray}
This vanishing of $g_5^{(2)}(p^2)$ is verified to a good accuracy on the lattice. Moreover, the \underline {pseudoscalar vertex at $q=0$ is entirely determined} from the mass function and $Z_{\psi}(p^2)$, i. e. from the propagator, through eq.~(\ref{Award3}), if we know the proportionality constant $m_{\rm R} Z_P$. 
However, this is not a practical way to determine the pseudoscalar vertex on the lattice, because of the very large Wilson artefact in $m(p^2)$. Rather, as we propose, this relation should be used in the reverse way~: to determine $m(p^2)$ from the pseudoscalar vertex and $Z_{\psi}(p^2)$. 

The proportionality constant $m_{\rm R} Z_P$ is obtained in a familiar way~; we define a bare mass $\rho$ through the equation $\partial_{\mu} j_A^{\mu}=2 \rho P_5$, which we can easily measure on the lattice through \underline {bare} matrix elements involving these operators (in the technical lattice practice, we rather denote by $\rho$ the same quantity in lattice units). Comparing this equation with the renormalised one  $\partial_{\mu} (j_A)_{\rm R}^{\mu}=2 m_{\rm R} (P_5)_{\rm R}$, we get at once~:  
\begin{eqnarray} \label{rho}
m_{\rm R} Z_P=Z_A \rho.
\end{eqnarray}
The good point in this transformation is that the r.h.s is independent of the renormalisation scheme, because it is so for $Z_A$ as we will show now.

Indeed, from eqs. (\ref{Award3}) and (\ref{rho}), one gets~:
\begin{eqnarray} \label{mp2}
m(p^2)= Z_A \rho~g_5^{(1)}(p^2)/Z_{\psi}(p^2)
\end{eqnarray}
Note that this relation has been demonstrated independently of any specific renormalisation scheme. Then $Z_A$ can be expressed as the same definite combination of bare quantities $Z_A=b(p^2)/\rho/g_5^{(1)}(p^2)$ in all schemes. However, this expression is not a practical way to determine $Z_A$ on the lattice, once again because of the very large Wilson artefact in the quantity $b(p^2)$ defined in eq.~(\ref{b}).

What is now to be noticed is that {\bf neither side of eq.~(\ref{mp2}) vanishes in the chiral limit if there is spontaneous breaking of the symmetry}. As to the $m(p^2)$ side, this is because the vacuum is then not invariant, therefore the expectation value of a scalar need not vanish, while on the other side there is a Goldstone pole in $g_5^{(1)}(p^2)$ compensating for the vanishing of the $\rho$ factor. We then write~:
\begin{eqnarray}\label{chir}
m_{chiral}(p^2)=\lim_{m \to 0} Z_A(m_q) \rho~g_5^{(1)}(p^2)/Z_{\psi}(p^2)= \nonumber \\
Z_A(m_q=0)~\lim_{m \to 0} \rho~g_5^{(1)}(p^2)/Z_{\psi}(p^2)
\end{eqnarray} 
This is the basic equation which we use below to deduce $m(p^2)$
in the chiral limit. We need both the functions $Z_{\psi}(p^2)$ and $g_5^{(1)}(p^2)$, which we determine from the measurement of the propagator and pseudoscalar vertex, as well as $\rho$, which will be measured through a ratio of vacuum expectation values as described below, and $Z_A(m=0)$, which will be taken from previous measurements by the ALPHA group (see also below).

{\bf An equation for the axial vertex}. Although not necessary for our direct purpose, we write for completeness the equation for the axial vertex parallel to the one relating the vector vertex to the quark propagator. Let us recall that one could use this relation to express $Z_V$ in terms of the bare propagator and vertex according to~:
\begin{eqnarray}
Z_V=Z_{\psi}(p^2)/g_V^{(1)}(p^2)
\end{eqnarray}
where $g_V^{(1)}(p^2)$ is the coefficient of the $\gamma_{\mu}$ term
in the Lorentz decomposition of the vertex. In practice, this relation is not the best suited to measure accurately $Z_V$ because of many artefacts. 

By taking the derivative of the axial W-T identity with respect to $q^{\mu}$, it is indeed similarly possible in principle to determine the axial vertex at $q=0$ from the propagator and the pseudoscalar vertex. 

In the axial case, the expression is more complicated that in the vector one; this is due to the   last term in eq.(\ref{AwardB}) which originates in the pseudoscalar density, and  reflects a non conservation of the axial current. What should be stressed is that once more the effect of this term does not vanish in the chiral limit if the symmetry is spontaneously broken. 

As in the vector current case, the W-T identity will give the constraints on the axial vertex by taking the derivative of eq. (\ref{AwardB}) with respect to $q$ at $q=0$. We get~:
\begin{eqnarray}\label{Award4}
Z_A \Gamma_{A\mu}= -i  \frac {\partial}{\partial p^{\mu}} S^{-1}(p) \gamma_5 + 2 i\, Z_A \rho \frac {\partial}{\partial p^{\mu}} \Gamma_5(p)
\end{eqnarray}
This relation again shows that $Z_A$ is independent of the renormalisation scheme. Of course, once more, this will hold up to terms vanishing as inverse powers of the cutoff at infinite cutoff, which are called {\it artefacts} in the lattice language. It must be recalled that, on the lattice, the Ward identity
is not exact, but holds only up to artefacts, because we work at finite cutoff, and the deviation will be found very large in some cases. Although theoretically possible, in practice, it is not easy to determine accurately $Z_A$ from it. This is why we recourse to other determinations from the ALPHA group.

\subsection{MOM renormalisation constants} \label{MOM} \hspace{\parindent}Useful and very usual specific renormalisation conditions for the Green functions on the lattice are the MOM ones, considered at some normalisation momentum
$p^2=\mu^2$, originally due to Georgi and Politzer. Although they are not really needed for our purpose, it remains useful to explain the connection with what precedes, since results are often discussed in terms of the renormalisation constants corresponding to this scheme. We lay particular emphasis on the need to account for Ward-Takahashi (W-T) identities in handling renormalisation. We start from the propagator and set~:
\begin{eqnarray}
S_{\rm R}^{-1}(\mu) = \delta_{a,b} \left( i\,\pslash + m_{\rm R}\right) \vert_{p^2=\mu^2},
\end{eqnarray} 
which means $Z_2^{MOM}=Z_{\psi}(\mu^2)^{-1}$ according to eq.~(\ref{Sm1}). Also, it means that the renormalised mass is then $m_{\rm R}^{MOM}=m(\mu^2)$, i.e. it is the mass function at $p^2=\mu^2$. This is the scheme of Georgi and Politzer.
 
For the vertices, we could think of choosing also the  standard MOM ones, i.e. tree level expressions for $p^2=\mu^2$, namely~:
  \begin{eqnarray}
(g_A^{(1)})^R(\mu^2) =1 \nonumber \\
(g_5^{(1)})^R(\mu^2)= 1
\end{eqnarray} 
This would be satisfactory at zero mass in perturbation.
However, we must recall that one is not free of choosing the renormalisation
of vertices once the scheme has been chosen for the propagator. Indeed, the renormalised theory must obey the symmetries, and this fact translates itself into
W-T identities strongly constraining the vertices. 

In the case of the vector current, the W-T constraint still implies that~:
\begin{eqnarray}
Z_V=\frac {Z_{\psi}(\mu^2)}{g_V^{(1)}(\mu^2)}=Z_V^{MOM}(\mu^2),
\end{eqnarray}
where $Z_V^{MOM}$ is defined by the standard MOM renormalisation condition~:
\begin{eqnarray}
(g_V^{(1)})^R(\mu^2) =1.
\end{eqnarray} 
But  this {\bf does not work for the axial current}. In effect, it is necessary to deduce the renormalisation of the axial vertex  from the W-T identities in a non perturbative treatment, with spontaneous symmetry breaking. From the W-T identities, one deduces that $Z_A \neq Z_A^{MOM}$, where $Z_A^{MOM}$ would be defined, in parallel with $Z_V^{MOM}$, as $Z_{\psi}(\mu^2)/g_A^{(1)}(\mu^2)$~; in fact $Z_{\psi}(\mu^2)/g_A^{(1)}(\mu^2)$ is not even independent of $\mu^2$~; one can expect only that it reaches $Z_A$ at large $p$~; then, it would be perhaps better to discard this MOM definition, since it could be misleading.

On the reverse, $Z_P$ can be defined, consistently with the Ward identities, from the standard renormalisation condition, analogously to $Z_V$. 
Indeed, setting $m_{\rm R}^{MOM}=m(\mu^2)$ and making $p^2=\mu^2$ in eq.~(\ref{Award3}), one gets~: 
\begin{eqnarray} \label{ZP}
Z_P(\mu^2)= \frac {Z_{\psi}(\mu^2)}{g_5^{(1)}(\mu^2)}=Z_P^{MOM}(\mu^2), 
\end{eqnarray} 
showing that $Z_P(\mu^2)$ as deduced from the W-T identity and MOM conditions for the propagator is equal to the one standardly defined directly by the tree condition~:
\begin{eqnarray}
(g_5^{(1)})^R(\mu^2)= 1.
\end{eqnarray} 

\subsection{OPE in the chiral limit and the quark condensate} \label{OPE}
\hspace{\parindent}The quark propagator, like any Green function, can be described by OPE at large momenta. In the present case, one deals with a non gauge invariant Green function, which implies the potential presence of non gauge invariant local operators in the OPE. The coefficient of all the operators giving the leading power corrections have been calculated by Lavelle et al. \cite{lavelle1992}. A great simplification is obtained if one takes {\bf the chiral limit}. Then, for the scalar part of the propagator or of its inverse, or the quark mass function, the perturbative contribution vanishes, and the dominating contribution in the OPE
is the quark condensate one~; this is in agreement with the fact that both the quark mass function and the condensate vanish with restoration of chiral symmetry . We stress this exceptional situation where the OPE begins by a power correction. This could lead in principle to a complementary determination of the condensate, but as we shall see in section \ref{OPEprincipal}, the attempt fails. The relevant quantitative formulae are given in this latter section. We just recall the tree level formula proposed a long time ago by Politzer \cite{politzer}, which gives the general structure of the contribution~:
\begin{eqnarray}
m(p^2)\simeq -\frac{4 \pi} {3} \alpha_s(p) \frac {1}{p^2}<\bar\psi \psi>
\end{eqnarray}

\subsection{A calculation of the condensate through the pseudoscalar correlator}\label{WTIcondensat}
\hspace{\parindent}\hspace{\parindent}A very interesting identity has been considered some years ago in several works on the lattice, which leads to a possible calculation of the condensate, advantageous for our discussion below, subsection \ref{pion-condensate}. This identity holds exactly when chiral symmetry is preserved by the lattice regularisation, and it has been indeed written for overlap fermions , see references \cite{edwards9811030,giusti0110184}~:
\begin{eqnarray}
<\bar\psi \psi>=-\lim_{m \to 0}~m \int d^4 x <P_5(x) P_5(0)> .
\end{eqnarray}
as adapted to our specific case $N_f=0$. The quantities are defined with the ``rotated" quark fields, often introduced in overlap calculations to improve Green functions (see the above references). Note that for $m \neq 0$ both sides retain \underline{additive} divergences proportional to powers of the mass, although the strongest one ,$1/a^3$, is cancelled by the quark field rotation. On the l.h.s, these divergences are due
to the limit $x \to 0$ taken in the propagator. On the r.h.s., the integration may imply divergences from coinciding arguments of the two $P_5$ composite fields, in addition to the multiplicative ones
coming from each $P_5$ field, so that these additive divergences cancel between the two sides. In the chiral limit, we have not such divergences, since they are cancelled by powers of $m$. 
In renormalised form, using for the overlap action $Z_P=Z_S=Z_m^{-1}$, we get~:
\begin{eqnarray}
<\bar\psi \psi>_{\rm R}=-\lim_{m_{\rm R} \to 0}~(m_{\rm R}  \int d^4 x <(P_5)_{\rm R}(x) (P_5)_{\rm R}(0)>)
\end{eqnarray}
Now, in the {\bf renormalised form}, it should hold independently of any particular choice of action, therefore also with the clover action, up to artefacts. Taking duly into account the renormalisation factors, we end, with the bare $<P_5(x) P_5(0)>$ on the r.h.s. , with~:
\begin{eqnarray}
<\bar\psi \psi>_{\rm R}=-\lim_{m_{\rm R} \to 0}~(Z_S (Z_P/Z_S)^2  m  \int d^4 x <P_5(x) P_5(0)>)
\end{eqnarray}
The ratio $Z_P/Z_S$ is independent of the renormalisation scheme.  Only $Z_S$ depends on it. The bare $<P_5(x) P_5(0)>$ on the r.h.s. can now be taken from any action, including the clover one. 

This identity can then be used on the lattice to calculate the condensate rather directly from the pseudoscalar correlator, for instance in the ${\overline{MS}}$ scheme. Indeed, the r.h.s. can be calculated with only standard logarithmic multiplicative renormalisations. We need not extract the pion residue, although this can be done as in the quoted papers, ending on the GMOR relation. It is useful for our purpose to avoid the recourse to the GMOR relation, because we want to calculate the condensate in the absence of the Goldstone state. In fact, it can be noticed that this direct method is the one proposed a very long time ago by  Bochicchio et al., the Rome group, \cite{bochicchio}. It can be established starting from the standard axial W-T identity~:
\begin{eqnarray}
\partial_{\mu} <A_{\rm R}^{\mu}(x) (P_5)_{\rm R}>=2 m_\rR <(P_5)_\rR(x) (P_5)_\rR>+\delta(x)<\bar \psi \psi>_\rR
\end{eqnarray}\r
and integrating over $x$. Note that we duly postulate the Ward identity in renormalised form. It is valid in bare form only up to finite renormalisation factors $Z_P/Z_S$. This Ward identity has been first established and exploited by D.J. Broadhurst \cite{broadhurst}.

This method is advantageous also with respect to the one of calculating directly
$<\bar\psi \psi>$ through the propagator~: it seems to circumvent the problem of extracting the power divergences.



\section{Lattice calculations with the clover action} \label{lattice}

\subsection{The problem of the Wilson term circumvented by the study of the vertex}
\hspace{\parindent}Of course, versions of the lattice Dirac action have been devised to improve the chiral behavior of the Wilson action, like the overlap, domain wall or twisted fermions.
The advantage of the Sheikholeslami-Wohlert(SW) or clover fermions is that they are relatively easily handled for not too small masses, in contrast to these more sophisticated versions. 
 
To calculate $m(p^2)$, the simplest way would seem to extract it directly from the lattice propagator, by extracting the scalar part. But this is not practicable for the Wilson or clover action, due to the large magnitude of the Wilson term, which affects the scalar part. Indeed, it is of order ${\cal O} (a p^2)$, where $a$ is the lattice unit~; this term is purely an artefact, but it cannot be avoided~; not only it is large, but moreover it increases like $p^2$, while the real, continuum $m(p^2)$ is decreasing like $1/p^2$ in the chiral limit, and otherwise logarithmically. Let us remind that the clover action improves the on shell quantities from order ${\cal O}(a)$ down to ${\cal O}(a ^2)$\cite{heatlie}, but not the Green functions, and presents the same large Wilson term artefact.

Fortunately, the problem is circumvented by the study of the vertex  \cite{cudell9810058}, because, in this case, no such an embarrassing artefact is present. Indeed, as we shall see, the data on the pseudoscalar vertex, with only a proper treatment of hypercubic anisotropy, present the expected, roughly power-like, decreasing behavior, see Fig. \ref{mp-chir-artevol-IR} for the related $m(p^2) \propto Z_P(p^2)$. Moreover, the smallness of artefacts is guaranteed by the good superposition of the data at $6.0$ and $6.4$ with proper renormalisation. This weakness of artefacts is only an empirical fact devoid of explanation. One may suspect that it is connected with the ``amputation" of external lines, i.e. for instance the tree approximation is exactly $\gamma_5$~; nevertheless, this is not a sufficient reason, for the analogous vector vertex is found to be still spoiled by large artefacts \cite{boucaud0307026,boucaud0504017}.

The connection between $m(p^2)$ and the pseudoscalar vertex is given in eq.~(\ref{mp2}) of the previous section, up to artefacts.Thereafter, for simplicity of notation, we denote by $m(p^2)$ the combination given by the r.h.s. of this equation, which should be equal to the scalar part of the propagator in the continuum limit, but which is quite different on usual lattices. We introduce lattice units~:
\begin{eqnarray} \label{mp2latt}
m(p^2)=a^{-1}~Z_A \rho~g_5^{(1)}(p^2)/Z_{\psi}(p^2).
\end{eqnarray} 
$\rho$ is now the dimensionless bare axial quark mass~:
\begin{eqnarray}
a \partial_{\nu} j_A^{\nu}= 2 \rho P_5 
\end{eqnarray} 
with $j_A^{\nu}$ and $P_5$ the bare lattice local axial current and pseudoscalar density and $a$ is the lattice spacing.  
In practice, $\rho$ is determined in the standard way through the ratio of v.e.v.'s~: 
\begin{eqnarray}
\rho= 1/2 \frac {\Sigma_{\vec x }<a \partial_0 j_A^0 (\vec x,0) P_5(0)>}{\Sigma_{\vec x}<P_5(\vec x,0) P_5(0)>}
\end{eqnarray}
Hereafter, we will define $\kappa_c$ as the value of $\kappa$ at which $\rho$ vanishes.All the other factors are dimensionless ab initio.
$Z_A$ is the $\mu$ independent renormalisation of the local axial current.  It can be determined by  certain W-T identities among  current correlators \cite{maiani1986} or other methods with specified renormalisation conditions. The lattice definition of $Z_{\psi}(p^2)$ is, as previously~:
\begin{eqnarray} \label{latticeZpsi}
Z_{\psi}(p)=\frac{1} {i} \frac {1} {12} {\rm Tr}\left[\gamma_\mu \bar p_\mu S^{-1}(p) \right]/(\bar p)^2 
\end{eqnarray}
where $\bar p_\mu \equiv \frac 1 a \sin \left( a p_\mu \right)$. Equation (\ref{mp2latt}) is the well known formula which has been used classically to determine the renormalised quark masses on the lattice \cite{allton9406263} at short distance. But here, it is used in the {\bf non perturbative regime}
of spontaneous chiral symmetry breaking, and in the chiral limit
where $m_q=0$ . In order to avoid an increase of errors on the chiral extrapolation, and since we are interested only in the chiral limit of $m(p^2)$, we multiply
by $Z_A$ only after having taken the chiral limit of the remaining factors. Now, the $Z_A$'s  in the chiral limit are accurately known by the work of the Alpha group \cite{luscher9611015}, and we borrow their central  values $Z_A(\kappa_c)$ . 

\subsection{The treatment of the raw lattice data}
\hspace{\parindent}The basic data, i.e. the quark propagator in the various configurations, are the same as already used in \cite{boucaud0307026,boucaud0504017} to study $Z_{\psi}(p^2)$. We have at hand simulations at $N_f=0$ with the Wilson gauge action and the SW clover action in Landau (i.e. Lorentz) gauge, on a series of lattices given in order of decreasing physical volume~:  $6.0,24^4$ $L=12.2~{\mathrm GeV}^{-1}$~; $6.0,16^4$, $L=8.14~{\mathrm GeV}^{-1}$~; $6.4,24^4$, $L=6.56~{\mathrm GeV}^{-1}$~; $6.6,24^4$, $L=5.06~{\mathrm GeV}^{-1}$~; $6.4,16^4$, $L=4.37~{\mathrm GeV}^{-1}$~; $6.8,24^4$, $L=3.93~{\mathrm GeV}^{-1}$. We have considered the inversion of the Dirac operator at five kappa values in each case~; namely at $\beta=6.0$, we choose $\kappa$ ranging from $0.1310$ to $0.1346$, corresponding to a large range of quark masses, so as to allow a reasonable chiral extrapolation~; the values at the other $\beta$'s are chosen to correspond approximately to the same bare masses in physical units, i.e. in terms of $m_q=\frac{1} {2a} (1/\kappa-1/\kappa_c)$, $m_q=0.233,0.154,0.104,0.054,0.0324~{\mathrm GeV}$. 

We first calculate the product $\rho  g_5^{(1)}(p^2)/Z_{\psi}(p^2)$ at each $\kappa$, for a given $\beta$ and volume $16^4,24^4$, and we correct this quantity for the hypercubic artefacts according to the same method used for the propagator vector part $Z_{\psi}(p^2)$ (see discussion below). Then we take the chiral limit $\kappa \to \kappa_c$ according to eq.~(\ref{mp2latt}), to obtain the chiral limit of the quark mass function.  The chiral limit is obtained by a fit in function of $\kappa$. The factor $Z_A(\kappa_c)$, which has a trivial effect is introduced as an additional fixed factor, not affecting the $p$ dependence and the essential conclusions.
A three parameters fit in $\kappa$ is possible with five $\kappa$ values, and it gives a sizeably better fit than with two parameters, pointing to a significant curvature in $m_q$. The coefficient of the ${\cal O}(m_q^2)$ term is strongly negative.

\subsection{Discussion of artefacts} \hspace{\parindent}We have given a very detailed discussion of artefacts in our previous papers on the quark propagator \cite{boucaud0307026,boucaud0504017}. We address the reader to these papers. However, it happens that some aspects of the discussion are crucial here, so that we give a new discussion for the relevant points. 

\subsubsection{Discretisation artefacts} 
\hspace{\parindent}As to discretisation artefacts, let us recall that we can classify them into two categories according to their behaviour under $O(4)$~: either they are not invariant under $O(4)$, which corresponds to the hypercubic artefacts, and we may determine them by using the various orbits~; or they are invariant and they can be extracted by using various $\beta$'s. It happens that in the present case, contrarily to the vector part of the propagator, both hypercubic artefacts and $O(4)$ invariant artefacts seem small.

\begin{figure}[hbt]
\begin{center}
\leavevmode
\mbox{\epsfig{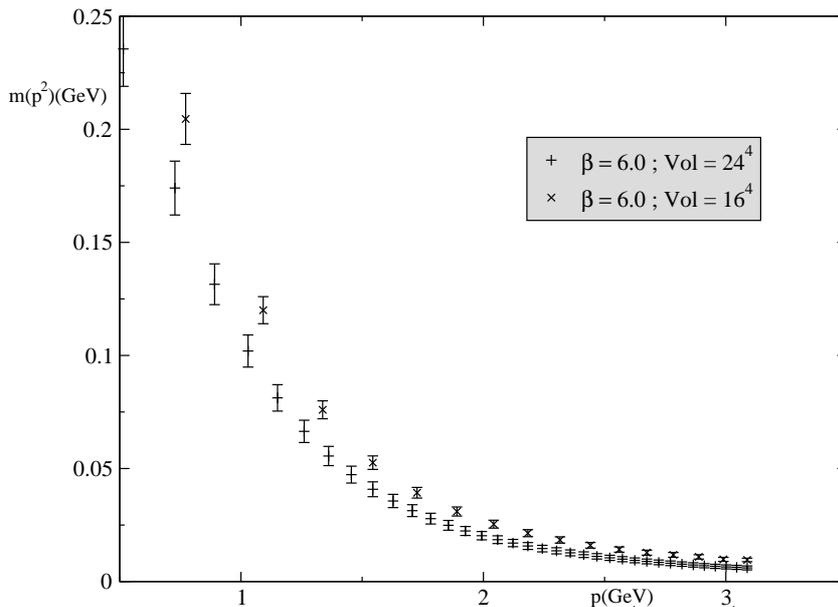}}
\vskip -0.5 cm
\caption{\small The chiral extrapolations at $\beta=6.0$, with $24^4$ and $16^4$ lattices, after elimination of hypercubic artefacts. A very regular behaviour is observed. The spacing between the two curves increases in the IR, which can be interpreted as a finite size effect.}
\label{mp-chir-artevol-IR}
\end{center}
\end{figure}

\begin{figure}[hbt]
\begin{center}
\leavevmode
\mbox{\epsfig{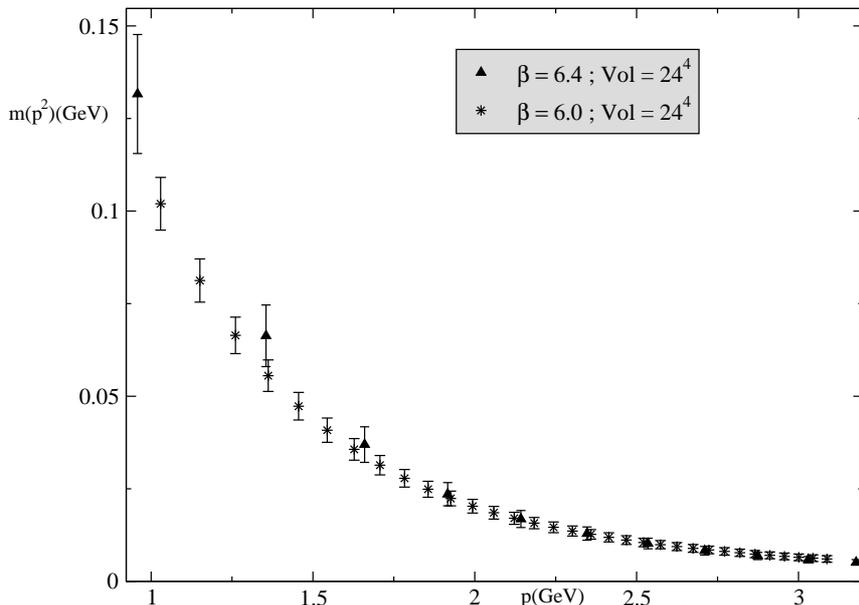}}
\vskip -0.5 cm
\caption{\small The chiral extrapolations at $\beta=6.0$ and $\beta=6.4$ at $24^4$ superpose very well on their common range of physical momenta. There remains however a small finite size effect.}
\label{mp-chir-6.0-6.4-IR}
\end{center}
\end{figure}

Note that we have treated the hypercubic artefacts by our systematic method of extrapolation, explained in several places, see for instance \cite{desoto07053523} \footnote{This technique was initially devised by C. Roiesnel}. With this method, we obtain data where almost any anisotropy has been eliminated, see Fig. \ref{mp-chir-artevol-IR} for an example, with the chiral extrapolation performed. 
\vskip 1cm
With such smoothed data, it is possible to perform very good analytical fits based on theoretical considerations in the continuum, with very low $\chi^2$. Possibly, the fit should also include terms accounting for $O(4)$ symmetric artefacts, to be determined by considering several $\beta$'s. In the present case, we even
do not require important terms of this sort. We have a good superposition of the chiral extrapolations at $6.0$
and $6.4$ with the $24^4$ lattice, on their common range of momenta, within statistical errors, see Fig \ref{mp-chir-6.0-6.4-IR}. However, a further, more detailed study, at large $p$, with multiplication by $p^2$, reveals the possibility of a small $O(4)$ symmetric artefact (see the end of the discussion of OPE, section \ref{OPEprincipal} and Fig. \ref{p2mp}). 
\vskip 1.5 cm

\begin{figure}[hbt]
\begin{center}
\leavevmode
\mbox{\epsfig{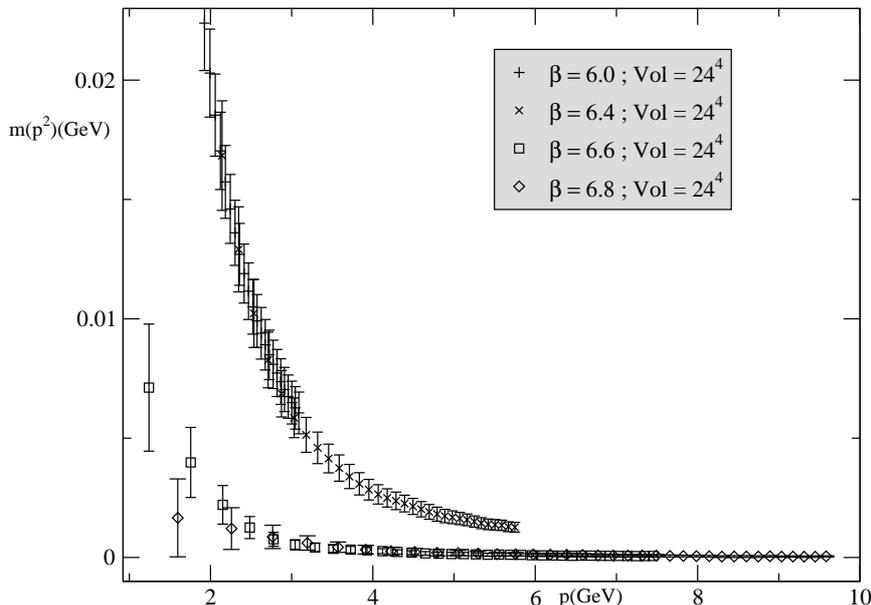}}
\vskip -0.5 cm
\caption{\small The chiral extrapolations at $\beta=6.0$ and $\beta=6.4$ superpose very well, with a relatively large value, corresponding to spontaneous chiral symmetry breaking. On the other hand, the chiral extrapolations at $\beta=6.6$ and $\beta=6.8$ also superpose very well, but with a zero value, corresponding to unbroken chiral symmetry. The huge difference between the two sets is a {\bf physical} volume effect, as explained in section \ref{discontinuity}.}
\label{mp-chir-UV}
\end{center}
\end{figure}

On the other hand, at $6.6$ and $6.8$ for the same lattice size, except for the first few points (where we have most probably a finite size artefact, see below), the chiral extrapolations superpose very well, but at a zero value, therefore completely different from the previous case, see Fig. \ref{mp-chir-UV}. Obviously, this is not a discretisation problem. We show in the next section that there is a discontinuity in function of the physical volume.
\vskip 0.5 cm

Having thus explained our own results, we give some comments on a recipe for treating discretisation artefacts , which leads to surprisingly different results in the case of the pseudoscalar vertex \cite{lubicz0403044}, {\bf in the chiral limit}. In addition to an usual democratic selection, reference \cite{lubicz0403044} reads the continuum $p_{\mu}$ as corresponding, on the lattice, not to the lattice $p_{\mu}$, but rather to  a trigonometric expression, $sin(ap_{\mu})/a$, differing from  $p_{\mu}$ by 
${\cal O}(a^2)$ terms. As far as the residue of the pseudoscalar vertex is concerned, our systematic method for eliminating hypercubic artefacts happens to give a  result close to the democratic selection if one reads $p^2$ in the democratic method as the $\Sigma_{\mu} p_{\mu}^2$ of the lattice. However a {\bf large discrepancy} appears when  \cite{lubicz0403044} identifies $p^2$ to $\Sigma_{\mu} (sin(ap_{\mu})/a)^2$, for the very asymmetric lattices used there (for example, $16^3 \times 52$ or $24^3,64$). Why the effect can be large can be easily understood~: the residue behaves roughly as $1/p^2$ in physical terms, and then, with the sine recipe, the curve appears much {\bf lower} at large $p$, because the ratio of the two curves is roughly $\Sigma (sin(ap_{\mu})/a)^2/\Sigma_{\mu} p_{\mu}^2$~; one finds differences as large as $50 \%$, for some asymmetric lattices at the largest momenta. Then, at large $p$, the recipe gives a notably smaller result than we find. Now, the question is~: what is the correct answer ?   

Our method gives a definite answer, by exploiting various orbits. As to the recipe, it could seem to be somewhat justified in some particular places, for free quarks, by comparing the continuum and lattice explicit expressions of the Green function~: for example, the vector part of the inverse quark propagator, which is found to be $i \Sigma \gamma_{\mu}~ sin(ap_{\mu})/a$ on the lattice, instead of $i \Sigma \gamma_{\mu}~ p_{\mu}$ in the continuum. Nevertheless, this has been taken into account by the standard lattice definition of $Z_{\psi}(p^2)$, eq.~(\ref{latticeZpsi}), which gives exactly one for free quark, like in the continuum~; now, when one considers not the free Green functions, but the non trivial $p^2$ dependence of the invariants like $Z_{\psi}(p^2)$ or $g_5^{(1)}(p^2)$ {\bf due to the interaction}, the recipe has no theoretical justification, and one can even doubt that there is a universal empirical recipe to reduce UV artefacts \footnote{The Adelaide group has indeed observed, in the case of the overlap action, and studying the scaling behaviour, that two different such recipes should be used for the vector and scalar part of the propagator \cite{zhang0301018}}.

\subsubsection{Lattice finite size artefacts} \label{finite-size}

\hspace{\parindent}The last type of artefacts is what will be usually termed as finite size
{\bf artefacts}, affecting the first few points in momentum space, i.e. the one with lowest momentum {\bf number}. Let us stress that it is quite distinct from the effect of the {\bf physical} volume, which we study in the next section, and which extends all over the range of available momenta and shows a discontinuity.

\begin{figure}[hbt]
\begin{center}
\leavevmode
\mbox{\epsfig{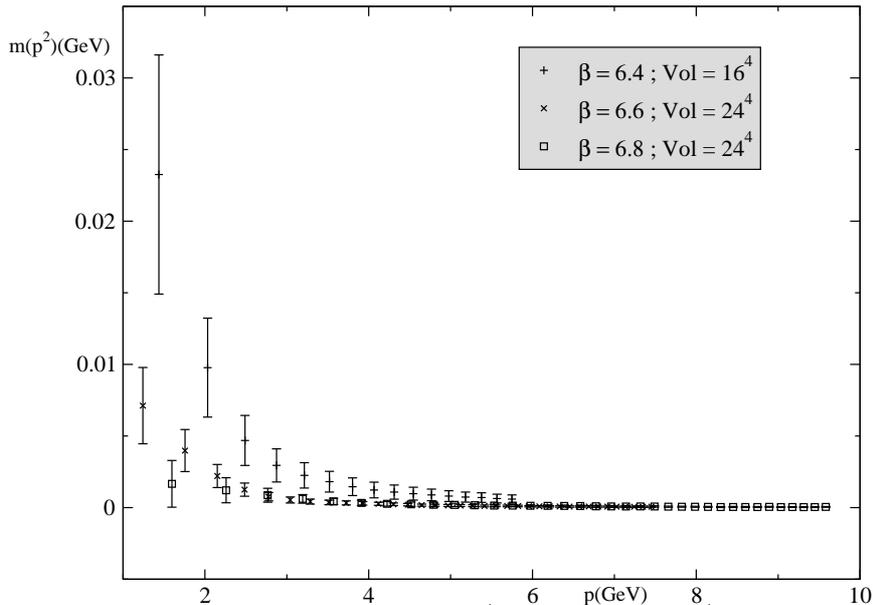}}
\vskip -0.5 cm
\caption{\small The cases $\beta=6.6$ or $\beta=6.8$ at $24^4$, and $\beta=6.4$ at $16^4$, corresponding to the restoration of chiral symmetry. One sees IR finite size effects, especially large in the latter case.}
\label{mp-chir-artevol-UV}
\end{center}
\end{figure}

As to these finite size artefacts, we find that they are  present. In fact, we observe that whatever the number of sites and the lattice unit, the first five points are always enhanced. We show this by considering first the cases where $m(p^2)$ is found to be very close to zero, Fig. \ref{mp-chir-artevol-UV}~: although the $\beta$'s are different, we can see that the value is extremely small ($10^{-4}~{\mathrm GeV}$ !!!) for $6.6$ and $6.8$ except for the first five points~; in this case, they differ from each other, which shows that it is an artefact. The data at $16^4$, which are larger, although still very small ($10^{-3}~{\mathrm GeV}$ !!!) except at the three first points, suggest the same interpretation, but also suggest that the finite size effects are notably larger at $16^4$. 
\vskip 0.5 cm
Then we observe that there is a similar effect in the other case $m(p^2)\neq 0$, by comparing the $24^4$ and the $16^4$ lattice at the same $6.0$, Fig. \ref{mp-chir-artevol-IR}, or $\beta=6.0$ and $\beta=6.4$ with size $24^4$, Fig. \ref{mp-chir-6.0-6.4-IR}.

\section{Discontinuity of chiral extrapolations in function of the physical volume} \label{discontinuity}

\subsection{Discontinuity of the quark mass function}

\hspace{\parindent}We observe two very distinct cases, already seen in Fig. \ref{mp-chir-UV} for the $24^4$ lattices~:

a)  at $\beta=6.0$ and $24^4$ or $16^4$, as well as at $\beta=6.4$ and $24^4$,
we obtain a non zero chiral limit, which superposes very well where possible, in the three cases, with large values at small $p$,
i.e. of order of $200~MeV$ at $p \simeq 0.5~{\mathrm GeV}$. This typically exhibits the behavior of a {\bf spontaneous breaking} of chiral symmetry.

The presence of a curvature in function of $m_q$ seems to explain why our chiral limit is smaller than in a previous calculation \cite{cudell9810058}, where only three $\kappa$'s were available~: for illustration we were getting around $m(2~GeV) \simeq 0.034~{\mathrm GeV}$, while we obtain now  $m(2~GeV) \simeq 0.022~{\mathrm GeV}$.

b)  at $\beta=6.6$ and $24^4$, and at $\beta=6.8$ and $24^4$, as well as at $\beta=6.4$ and $16^4$, we obtain on the contrary much smaller values on the whole range of momenta~; moreover, if we recalculate $\kappa_c$ by defining it through $\rho(\kappa_c)=0$ in each corresponding finite volume, it is a bit different from the standard one determined in very large volumes, and then the chiral limit is {\bf very small} and physically not significant \footnote{In view of this observation, we have recalculated also the case a) with the same prescription for $\kappa_c$.}, except for the low momenta $p<3~{\mathrm GeV}$ of the $6.4,16^4$ lattice, where there is a value significantly different from zero, although small, and monotonously decreasing with $p$. For these latter points, the most natural explanation seems a volume artefact on the four or five points, which is also present in the other cases, with a different magnitude, as we have argued.
On the whole, the case b) seems typically a situation of {\bf restoration of chiral symmetry}.

Now, the remarkable fact is that chiral symmetry ``seems" to be ``restored" rather abruptly
for volumes smaller than a certain {\bf physical} value $L_c  \simeq 6~{\mathrm GeV}^{-1}$, 
in the sense that standard chiral extrapolation by a low polynomial in $m_q$ yields $m_{chiral}(p^2) \neq 0$ above $L=L_c$ and $m_{chiral}(p^2)=0$ below, where $m_{chiral}$ denotes the chiral limit. Indeed, the relevant parameter distinguishing case a) and case b) seems to be the {\bf physical} volume. The physical length of the lattice is respectively $L=12.2, 8.13, 6.55~{\mathrm GeV}^{-1}$ for the case a), and $L=5.05, 4.37, 3.93~{\mathrm GeV}^{-1}$ for the case b), the order being the same as above. The separating length is then around $L_c \simeq 5.8~{\mathrm GeV}^{-1}=1.2~fm$ (we choose the middle between the two lengths). Moreover, we speak of a {\bf discontinuity} and not simply of a transition~: it is because  the last volume which presents ``symmetry breaking"  is certainly larger than the first one which presents ``symmetry restoration", but the two volumes are not very different~: $\beta=6.4, 24^4,L=6.56~{\mathrm GeV}^{-1}$ against $\beta=6.6, 24^4,L=5.06~{\mathrm GeV}^{-1}$. Moreover, on each  side of the  discontinuity, the results for the order parameter $m_{chiral(p^2)}$  are very similar for the three lattices of case a), and very similar for the three lattices of case b). In the case of spontaneous breaking, this is especially striking, because the volumes extend over a large range~; the largest volume, $\beta=6.0, 24^4$ can be considered as a relatively large volume, $V=2.2~10^4~{\mathrm GeV}^{-4}$ but the smallest one $\beta=6.4, 24^4$ which is ten times smaller, $V=1.85~10^3~{\mathrm GeV}^{-4}$~; and the latter volume is much closer to the $\beta=6.6, 24^4$ volume $V=6.55~10^2~{\mathrm GeV}^{-4}$ for which symmetry is manifestly restored. This means a discontinuity around $V=1.25~10^3~{\mathrm GeV}^{-4}$ (middle point) or, in length, around $5.8~{\mathrm GeV}^{-1}$.

In \cite{boucaud0504017}, we studied the chiral symmetry breaking through the overlap $\Gamma_A/\Gamma_V$ as a possible indicator of spontaneous symmetry breaking~: it should differ from 1 due a Goldstone contribution at small $p$~; in fact, we saw an effect at small $p$ presumably coming from the Goldstone, at least at $\beta=6.0$. However, at small volumes corresponding to $6.6,6.8$, the available momenta were too large to test whether this Goldstone effect disappears or not. Now, the study of the pseudoscalar vertex (or quark mass function), for which the Goldstone effect is much larger when present, gives a clear answer~: it shows a striking vanishing of the Goldstone effect at $6.6, 6.8$.

A typical feature is that this transition affects $m_{chiral}(p^2)$ over the whole momentum range  simultaneously, i.e. when passing the same critical length. Of course, there are also, as we have shown in subsubsection \ref{finite-size}, finite size artefacts which somewhat enhance the smallest momenta. But they superpose on top of a very clear discontinuity with respect to the physical volume. And, even where enhanced by this volume artefact,i.e. at small momenta, the curves of case b) lie much lower than those of case a)~: admittedly, the $6.4,16^4$ is also lower, but less than the others at the very first few points.

To reinforce our conviction that we are indeed facing a notable phenomenon,
we propose in the next subsection two other similar and striking observations, which display the same critical length.

\subsection{Discontinuity of  the pion mass and of the condensate value} \label{pion-condensate}

\hspace{\parindent}We extract these two observations from the behaviour of the pseudoscalar correlator, which we have calculated together with the vertex on the same  lattices  with the same four $\beta$ values. From it, we can calculate,
  
\begin{minipage}{12cm}
i)  the pion mass, according to the well known method using the $<P_5(x) P_5(0)>$ correlator integrated over space,  at large $t$,
\end{minipage}

\noindent but also 

\begin{minipage}{12cm}
ii)  the quark condensate, through the method explained above (subsection \ref{WTIcondensat}), i.e. through the chiral extrapolation of $-m \int d^4 x <P_5(x) P_5(0)>$. 
\end{minipage}

\noindent For both quantities, on performing a low order polynomial extrapolation in $m_q=\frac {1} {1/2a} (1/\kappa-1/\kappa_c)$, we observe exactly the same type of discontinuity as for $m(p^2)$, i.e. everything goes as if one had a phase transition at small physical volume. The results are displayed in the table below and graphically in figure \ref{restore-symmetry}.

\begin{center}
\begin{tabular}{|l|c|c|c|c|}
\cline{1-4}
\multirow{2}{6mm}{$\beta$}&Size (Length)&$m_\pi^2$&\raisebox{-.4mm}{$\langle\overline{\psi}\psi\rangle$}\\
&\phantom{Size }( GeV$^{-1}$)&({GeV}$^{2})$&(GeV$^{3})$\\
\hline
\multirow{2}{6mm}{6.0}&$24^4$ (12.2)&$(1.54\pm1.58) 10^{-2}$&$(-44.3\pm7.4) 10^{-3}$&\multirow{3}{15mm}{a) ``broken symmetry"}\\\cline{2-4}
&$16^4$ (8.14)&$(4.05\pm2.59) {10^{-2}}^*$&$(-50.7\pm7.9) 10^{-3}$&\\
\cline{1-4}
6.4&$24^4$ (6.56)&$(-5.1\pm 4.6) 10^{-2}$&$(-43.5\pm7.8) 10^{-3}$&\\
\hline
\hline
6.6&$24^4$ (5.06)&$(0.66\pm 0.115)$&$(-7.29\pm14.3) 10^{-3}$&\multirow{3}{15mm}{b) ``restored symmetry"}\\
\cline{1-4}
6.4&$16^4$ (4.37)&$(0.615\pm0.21) $&$(-24.5\pm12.1) {10^{-3}}^{*}$&\\
\cline{1-4}
6.8&$24^4$ (3.93)&$(1.3\pm 0.22)$&$(-8.65\pm28.8) 10^{-3}$&\\
\hline
\end{tabular}
\end{center}

We draw the following conclusion from this table~:

i)  the chiral extrapolation gives a much smaller $m_{\pi}^2$ in case a) than in case b), by two orders of magnitude. This striking difference confirms the advocated discontinuity. In fact, it is almost compatible with $0$, $m_{\pi}^2 \simeq 0$, in case a)~- and this is in agreement with spontaneous symmetry breaking. For the largest volume, it is fully compatible with $0$. On the other hand, there is no Goldstone boson in case b), but a heavy meson. A complementary criterion for the study of chiral restoration would be the presence of a scalar meson degenerate with the pseudoscalar one.

ii)  the chiral extrapolation of $m \int d^4 x <P_5(x) P_5(0)>$ gives something close to the expected value for the quark condensate, or larger than it for the case a).

The central values of the corresponding renormalized condensates in ${\overline{MS}}$ at $2~{\mathrm GeV}$ are estimated to be, either by using values of $Z_S^{RI-MOM}$  non perturbatively measured or those obtained through $Z_P^{RI-MOM}$, and using also $Z_P/Z_S$ from Ward identities~:$<\bar{\psi} \psi>_{\overline{MS}}(2~GeV) =-(2.2 \pm 0.4)~ 10^{-2}~{\mathrm GeV}^3$ at $6.0, 24^4$, $<\bar{\psi} \psi>_{\overline{MS}}(2~GeV) =- (2.5 \pm 0.4)~ 10^{-2}~{\mathrm GeV}^3$ at $6.0, 16^4$, in a quite encouraging agreement with standard values from the GMOR relation at $N_f=0$. Indeed, from \cite{lubicz0403044}~: 
\begin{eqnarray} \label{GMOR-lubicz}
<\bar{\psi} \psi>_{\overline{MS}}(2~GeV)=-(273 \pm 19~MeV)^3=(2.0 \pm 0.5)~ 10^{-2}~{\mathrm GeV}^3 \end{eqnarray}
On the other hand, the renormalised value $<\bar{\psi} \psi>_{\overline{MS}}(2~GeV) =-(3.2  \pm 0.6)~10^{-2}~{\mathrm GeV}^3$ obtained at $6.4,24^4$ is somewhat too large, but not far from compatibility with eq.~(\ref{GMOR-lubicz}).

On the contrary, for the case b), it is smaller, and compatible with zero for two lattice out of three. The exception is the $6.4,16^4$, which gives an unexpected intermediate value. So the conclusion would be equally striking , if not for the exception of this small volume lattice $6.4, 16^4$. 

This exception does not correspond to an intermediate physical volume (this would invalidate our advocated conclusion of a discontinuity as function of the physical volume). Rather, it is probably due to a finite size artefact, connected with the similar observation for $m(p^2)$. At the same $6.4,16^4$, $m(p^2)$ shows rather high points at the first momenta, much higher than those for the others in case b). Such finite size artefacts seem also to affect the case a)~; the renormalised value of the condensate at $6.0,16^4$ is somewhat larger that the one at the largest volume $6.0,24^4$~; simultaneously, the pion mass is somewhat larger. It is difficult to give a complete rational explanation of the size of such artefacts. Nevertheless, it is doubtless that finite size artefacts
are superimposed on the basic physical volume effect, and can be distinguished from it because they do not follow the same {\bf rationale}.

\subsection{Comments on the unexpected discontinuity of chiral extrapolations in function of the {\bf physical} volume} \label{chiS}

\subsubsection{A phase transition ?}
\hspace{\parindent} At this point, it is important to enter into some warnings, to avoid possible confusions. 

i) What is as expected. We are well aware that chiral symmetry is not expected to undergo spontaneous  breakdown in finite volumes~: order parameters are expected to vanish anyway when $m_q \to 0$. But, {\bf if the volume is sufficiently large}, this may happen at only very small quark masses, which we cannot consider with Wilson-type actions. Then, at moderate masses reached by Wilson-type actions, things, as is well known, may look as in the infinite volume case~: order parameters like the mass function seem to tend to the same non zero value as in the infinite volume, if one performs a naive chiral extrapolation with a low order polynomial in $m_q$. Of course, $m(p^2)$ as function of $m_q$ should present a bending downwards with respect to this extrapolation, if one were able to go to smaller masses. To know whether the volume is sufficiently large, a necessary condition is to check whether the ``chiral" limit is stable against variation of the volume, which we verify.  Let us call this first situation ``infinite-volume-like". It seems to correspond to our case a).

On the other hand, if the volume is sufficiently small, it is natural to expect
that, simply extrapolating from the moderate masses, one gets already a vanishing of order parameters. Let us denote this second situation as the one of  ``chiral symmetry restoration". It seems to correspond to our case b). 

ii) What is surprising. This dichotomic presentation is however an oversimplification, according to the common ideas. One would expect
a {\bf continuous transition} between the two situations when one decreases the volume~; the chiral extrapolation would be expected to deviate progressively from the truly infinite volume value, and decrease downwards to zero. In the same vein, the Goldstone boson would be expected to acquire progressively a mass.  Let us recall for instance the finding in the so called $\epsilon$ regime for the condensate~: the ratio to the infinite volume limit deviates from 1 by a function of 
$z=m~V$, therefore,  at fixed mass, it is a continuous function of the volume 

In face of such expectations, our lattice analysi show on the contrary an abrupt discontinuity between the two cases a) and b) in a rather narrow window of physical lattice length or volume. Indeed, our low order polynomial extrapolations exhibits this  striking {\bf discontinuity} for $m_{chiral}(p^2)$, which drops suddenly to zero.  And the Goldstone boson acquires abruptly a very large mass. This discontinuity would suggest speaking of a phase transition. In fact, this would be a too strong statement, since principles seem to be against such a conclusion for finite volumes, and also, since we are not really experimenting the chiral limit, with truly very small masses, but instead performing only an {\bf extrapolation}. Rather, one should speak of a sharp transition to a new regime of chiral extrapolation~: 

-above the critical length, the naive chiral extrapolation picks the quantities corresponding to  infinite volume and  spontaneous  breaking,

-while thereunder it picks the ones corresponding to chiral symmetry restoration.

\begin{figure}[hbt]
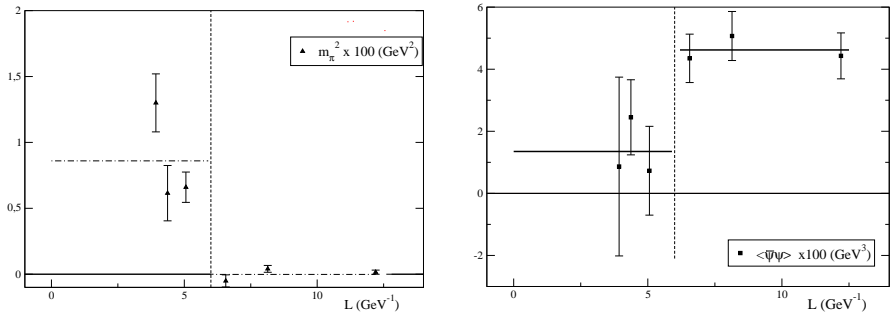
\begin{center}
\begin{tabular}{c c}
\includegraphics[width=5.5cm]{mpi2.eps}\hspace{.2cm} &\includegraphics[width=5.5cm]{condseulnoir.eps}\\
\end{tabular}
\caption{\small $m_\pi^2$ (left) and the quark condensate (right) as functions of the lattice size. The leftmost part of each graph shows the restoration of chiral symmetry at small volume while the symmetry is broken in the rightmost one.  The horizontal lines are indicative of the average value of the measured quantity in each region. Note that for $m_\pi^2$ in the symmetry breaking phase the errors are very small and can hardly be seen on the figure.}\label{restore-symmetry}\end{center}
\end{figure}

\noindent This behaviour is illustrated in figure \ref{restore-symmetry}

\subsubsection{Origin of the discontinuity}

Our observation deserves obviously  understanding. We are tempted to assume a connection with the observation of Neuberger and Narayanan\cite{narayanan} of a restoration of chiral symmetry below some critical length. Admittedly, they work in the $N_C \to \infty$ limit and they then expect a true phase transition at finite volume. However, their critical length is close to ours~: they find $L_{crit}$ around $1~fm=5~{\mathrm GeV}^{-1}$.
 
 \vspace{4mm}
 
One can also think of a connection with the ``finite temperature transition", which affects both the confining properties and the chiral symmetry and is observed at large spatial volume and finite time interval $1/T$. Indeed the transition occurs around $T_c=0.270~{\mathrm GeV}$, therefore $1/T_c \simeq 3.7~{\mathrm GeV}^{-1}$, mot too far from our $L_c=5.8~{\mathrm GeV}^{-1}$.  However, our data do not correspond to this situation of highly asymmetric lattices which would allow to study such questions.

Let us emphasize that the transition we observe, concerning chiral symmetry, is not a universal fact. Not all condensates vanish at small volumes. In our analyses of the gluon, ghost and vector part of the quark propagator, the non perturbative v.e.v $<A^2>$ was found the same, consistently, at all the four volumes $\beta=6.0,~6.4,~6.6,~6.8,~24^4$, the same volumes as we use now. $<A^2>$, which does not seem to be an order parameter for some symmetry breaking, does not collapse in small volumes. Note that we are not yet in a situation of such a small volume that everything should be  perturbative simply because power corrections would be negligible (``femto-universe").

\section{The OPE} \label{OPEprincipal}

\subsection{Failure of OPE with the Wilson coefficients at low order} \hspace{\parindent}Let us now concentrate on the case  a), i.e. the one where  the chiral limit mass function is large. From now on, we shall consider only renormalised quantities, in the ${\overline{MS}}$ scheme, and skip any subindex meant to recall the renormalisation~: we quote only the subtraction point. Let us first recall the puzzle underlined in \cite{cudell9810058}, where one studies the results of lattice data by the QCDSF group at $\beta=6.0$.  If we consider the OPE with tree level coefficient for the condensate,
we have in the chiral limit the Politzer formula quoted above \cite{politzer}~:
\begin{eqnarray}
m(p^2)= -\frac{4 \pi} {3} \alpha_s(p) \frac {1}{p^2}<\bar\psi \psi>
\end{eqnarray}
Note the remarkable fact that the expansion begins with the power correction. The purely perturbative contribution vanishes in the chiral limit, since it is proportional $m_q$. 

We observe immediately an {\bf enormous discrepancy} between both sides at the momenta usually considered~; indeed, with the estimate of  $<\bar\psi\psi>=-[(0.267 \pm 5 \pm 15)~$GeV$]^3=-(1.9 \pm 0.35)10^{-2}~$GeV$^3$ from the mass of the pion and GMOR formula with overlap fermions
(\cite{giusti0110184}), or $-(2. \pm 0.5)10^{-2}~$ GeV$^3$ from clover fermions, as given above \cite{lubicz0403044}, see eq.~(\ref{GMOR-lubicz})~; taking $p=2~GeV,\alpha_s(p) \simeq 0.3$ (at one loop, with $\Lambda_{\overline{MS}}=0.240~{\mathrm GeV}$), we find a r.h.s. around $5~10^{-3}~{\mathrm GeV}$, much smaller, by a factor 4, than the value we find from the lattice result for the l.h.s.~: $m_{latt}(2~GeV) \simeq 2~10^{-2}~$GeV .Note that there is an habit to present condensates through cubic roots, which hides the discrepancies. We want to avoid it. Le us recall that $4^{1/3}\simeq 1.6$. Note also that, in the initial paper \cite{cudell9810058}, the discrepancy was found still larger, around 10. The reason is twofold~: first we adopted as reference the  standard QCD sum rule value of the condensate, $<\bar\psi\psi>=-(0.225~GeV)^3=-0.0114~{\mathrm GeV}^3$~,which is twice  smaller ; second, the lattice value of $m(p^2)$ that we estimated was larger, as explained in the beginning of section \ref{discontinuity}, because we could do only a linear extrapolation. Nevertheless, the discrepancy remains huge.

Working at higher momentum would not help much in this respect~: using the $\beta=6.4,24^4$ lattice, we have momenta up to more than $5~{\mathrm GeV}$, but nevertheless the discrepancy is not much smaller~: the prediction at tree level, with $\alpha_s(5.7~GeV)\simeq 0.18$ $m(5~GeV) \simeq  \frac{4 \pi} {3} 0.18 \frac {1}{5.7^2} 0.0176~GeV \simeq 4~ 10^{-4}~{\mathrm GeV}$, against $m_{latt}(5.7~GeV)= (1.15 \pm 0.67)~10^{-3}$ according to our lattice measurements, therefore there is a factor about three of discrepancy. Note that here we choose to calculate   $\alpha_s(p)$ at one loop for simplicity in this first discussion. To be more quantitative, it seems that we should adopt the best possible approximation for $\alpha_s$, i.e. including $\beta_3$. \underline{This is what we do in the rest of the section}. At $N_f=0$, the resulting values of $\alpha_s$ are sizeably smaller, which reinforces the problem. 

We  can include, as done already in \cite{cudell9810058}, the one-loop correction to the coefficient, which has been
calculated a long time before by Pascual and de Rafael \cite{pascual1981}~; we set in their formula $a=0$ for the Lorentz gauge, $N_c=3$, and for the renormalisation point, $\mu=p$ in their notation~; we also take their $p^2$ as minus the Euclidean $p^2$ of the lattice~; whence~:
\begin{eqnarray}
m(p^2)= -\frac{4 \pi} {3} \alpha_s(p) \frac {1}{p^2}<\bar\psi \psi> (p)\left(1+6.1875 \frac {\alpha_s(p)} {\pi}\right) 
\end{eqnarray}
We can in addition make a fit on the whole range of our data. We take into account the evolution of the condensate from $p$ down to the reference point $2~{\mathrm GeV}$ where we want to determine the condensate, by using the formula for the evolution of the ${\overline{MS}}$ quark mass, which is just the inverse of the one for $<\bar\psi \psi> (p)$~:
\begin{eqnarray}
<\bar\psi \psi> (p)/<\bar\psi \psi> (2~GeV)=((\alpha_s(2~GeV) /\pi)^{4/11}\nonumber \\ (1+0.687328~(\alpha_s(2~GeV) /\pi)+ \nonumber \\
1.51211~(\alpha_s(2~GeV) /\pi)^2+4.05787~(\alpha_s(2~GeV) /\pi)^3))/\nonumber \\
((\alpha_s(p) /\pi)^{4/11}~(1+0.687328~(\alpha_s(p) /\pi)+ \nonumber \\
1.51211~(\alpha_s (p)/\pi)^2+4.05787~(\alpha_s(p) /\pi)^3))
\end{eqnarray}
to the corresponding order. We have checked this formula by calculating it from both the expressions in \cite{vermaseren} and \cite{chetyrkin9703278,chetyrkin0007088} respectively. It coincides exactly with the one given in \cite{chetyrkin09110594}.

In the common range of momenta at $\beta=6.0$ and $\beta=6.4,24^4$, $p=2-3~{\mathrm GeV}$, cf Fig. \ref{mp-chir-6.0-6.4-IR}, we find a good superposition of the lattice curves for $m(p^2)$ at $6.4$ and $6.0$, and a very consistent fit where the ``condensate" at $2~{\mathrm GeV}$ in ${\overline{MS}}$ scheme would have the fitted value $-(0.062 \pm 0.01)~{\mathrm GeV}^3$ at $\beta=6.0$, $-(0.056 \pm 0.009)~{\mathrm GeV}^3$ at $\beta=6.4$, therefore around
3 or 3.5 times the actual value. This is still clearly unacceptable. The situation is only slightly better if we use larger momenta, available only with the $6.4$ lattices~; in this case, using the full
range of momenta, we get $-(0.05 \pm 0.01)~{\mathrm GeV}^3$  \footnote{Note that even introducing a strong additional $1/p^4$ term and with a different treatment of discretisation effects, the condensate in \cite{lubicz0403044} remains sizeably larger than required~: $-(0.312)^3~{\mathrm GeV}^3 \simeq -0.03~{\mathrm GeV}^3$~; without the $1/p^4$ term, it rises to a still larger value, larger by a factor $(.792/0.721)^3=1.33$, whence $-0.04~{\mathrm GeV}^3$, i.e. a factor 2 of discrepancy.}.

Obviously, the discrepancy is so huge that we have to find a non trivial explanation. We can think at once of two types of explanation of the discrepancy, preserving the OPE~: -either the above condensate value, taken from the pion mass, has something basically wrong~; it is hard to believe~; note also that many successful considerations have been based on roughly this value, and one would have to reconsider a whole sector of particle physics~; -or it is the perturbative expansion of the coefficient which is strongly modified by higher orders, or even, which is simply not valid, at least at available momenta.   In \cite{cudell9810058}, the second explanation is suggested, i.e. it was suggested that the perturbative expansion may merely break down at available lattice momenta, since already the one-loop correction is found as large as $50\%$ of the tree level at $2~{\mathrm GeV}$. It could mean that either there is merely no sort of convergence or that the OPE is practically useless. Equivalently, it could mean that we have to reach very high momenta for the known, low order, perturbative expansion to be valid. However, since the decrease of $\alpha_s$ is very slow, we would require prohibitively large momenta to test the idea on the lattice. 

\subsection{The lesson from very high order calculations}

Fortunately, an impressive progress has been performed in the calculation Incorporating the higher order terms recently computed by K. Chetyrkin and A. Maier\footnote{We thank them for communicating their work to us prior to publication} gives (cf eq.~(\ref{ordre4}) in the appendix)~:

\begin{eqnarray}
&m(p^2)\simeq& 
-4 \pi/3 \alpha_s /p^2 <\bar \psi \psi>(p)\nonumber \\   
&&(1+6.1875 (\alpha_s /\pi) + 
52.9495 (\alpha_s /\pi)^2+
564.8284 (\alpha_s /\pi)^3) 
\end{eqnarray}
We write $\simeq$ because there is in addition a correction of the same formal order from the quark self-energy, $\Sigma(p)$, beginning at two loops, which is found to be very small($1 \%$). The positive and steadily increasing coefficients are much larger than in the purely perturbative series for the propagator(see for instance eq. \ref{sigma}). It do much in convincing one that the perturbative expansion of the Wilson coefficient is not well behaved. For $p=2{\mathrm GeV}, \alpha_s \simeq 0.2$ one gets~:

\begin{eqnarray} \label{ordre4num2}  
m((2~{\mathrm GeV})^2) \simeq 4 \pi/3~0.2/(2~GeV)^2~(-<\bar\psi \psi>(2~{\mathrm GeV}))\nonumber \\ (1+0.394+0.215+0.146)
\end{eqnarray}
On one hand, it shows that the series is meaningless at $2~{\mathrm GeV}$. And it implies that claims to recover the value of the condensate from data around $2~3{\mathrm GeV}$ with only the low orders cannot be justified. On the other hand, it suggests that the discrepancy
we have observed at lower orders between the fitted and the 
real value of $<\bar\psi \psi>$ could be explained by the strength of high order radiative corrections.
 
\begin{figure}[hbt]
\begin{center}
\leavevmode
\mbox{\epsfig{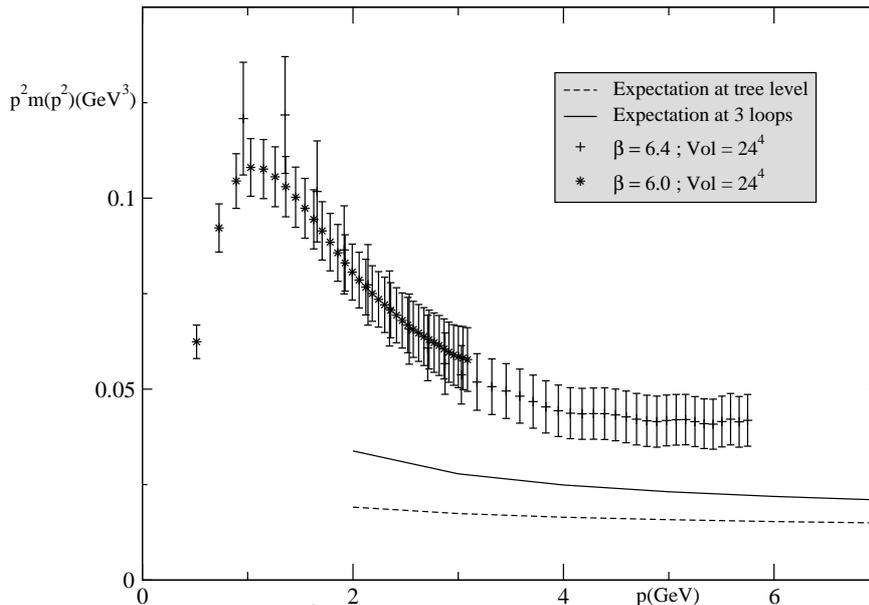}}
\vskip -0.5 cm
\caption{\small The lattice data $\times p^2$ at $6.0$ and $6.4$, compared with the OPE expectations at tree level and at three loops, with an upper value of the condensate ($2.25~10^{-2}~{\mathrm GeV}^3$) estimated by combination of overlap and clover estimates in the GMOR method, showing the gap between the lattice data and the OPE expectations.}
\label{p2mp}
\end{center}
\end{figure}

Working at higher $p$ would presumably not suffice, even assuming some convergence of the series. With $\alpha_s(5~{\mathrm GeV})/\pi \simeq 0.147/\pi$, the behaviour of the series is substantially improved~:
\begin{eqnarray} \label{ordre4num3}  
m((5~{\mathrm GeV})^2) \simeq -4 \pi/3~0.147 /(5~{\mathrm GeV})^2~(-<\bar\psi \psi> (5~{\mathrm GeV}))\nonumber \\
 (1+0.289+0.116+0.058)
\end{eqnarray}
and, as can be seen on figure (\ref{p2mp}) the inclusion of the higher terms results in reducing the gap between the lattice data and  the OPE estimation by a factor of 2. But is it is unable to reconcile them. Equivalently, the fit gives  a fictitious value for the condensate of   $-(4. \pm 0.6)10^{-2}~$ GeV$^3$, twice too large. To summarize, not even on our $6.4,24^4$ lattice, where momenta run up to more than $5~{\mathrm GeV}$ , can we  apply safely OPE.   In the case of  $Z_{\psi}(p^2)$ \cite{boucaud0504017}, we could use the $6.6,6.8$ lattices, which provide momenta up to $10~{\mathrm GeV}$, to study the $<A^2>$ condensate; this is not possible for the quark condensate in  the present case, because of the collapsing to zero of the quark mass function at those small volumes.

What is encouraging is that the terms of the series are all positive, and, if this behaviour persists, it makes understandable that the real curve lies above the OPE contribution presently calculated. 

\subsection{ Remaining discretisation artefacts ?}
\begin{figure}[hbt]
\begin{center}
\leavevmode
\mbox{\epsfig{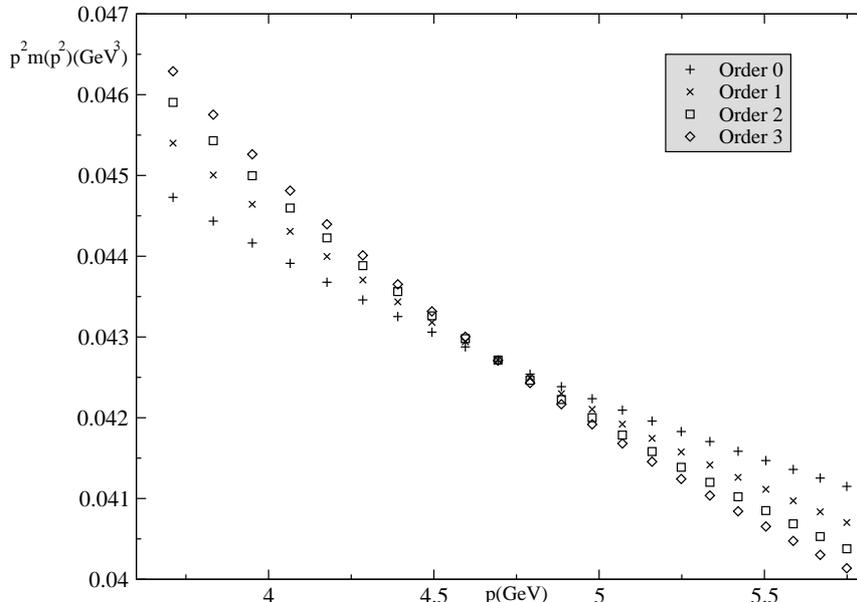}}
\vskip -0.5 cm
\caption{\small The fits of $p^2~m(p^2)$ with increasing number of loops in the Wilson coefficient of the condensate. The fits are performed over the largest momenta available (6.4), to favor as much as possible the convergence the OPE, showing the increasing, although  slowly evolving, slope.}
\label{p2fitp_6.4_UV}
\end{center}
\end{figure}

 Of course, one should consider the possibility that there remain discretisation artefacts. We feel that in spite of the good superposition of $6.0$ and $6.4$, there is some {\bf positive} $O(4)$ invariant artefact
(e.g. $+a^2p^2$), but presumably much too small to explain the discrepancy.
Signals of a small artefact (see Figure \ref{p2mp} are 1) $6.0$ and $6.4$ differ somewhat around the endpoint of $6.0$, $p \simeq 3~$GeV (see Fig. \ref{p2mp})~;  2) the central curve of the $6.4$ points seems too flat beyond $4~$GeV, although our errors are too large to ascertain this statement. About point 2)~: the theoretical perturbative curve is falling more and more rapidly with increasing order, as seen in Figure.\ref{p2fitp_6.4_UV}. One may then expect the exact continuum curve to fall still more rapidly. and not to be flat. The flatness of the lattice data could be explained by a positive artefact of the type $a~p^2$  or $a^2~p^2$, corresponding to an artefact $\delta m(p^2) \propto a$ or $a^2$. The magnitude of this artefact is however strongly limited by the region around $p \simeq 3~$GeV, where both $6.0,6.4$ points are present and give close central values of $p^2~m(p^2)$ (difference is around $0.4~10^{-2}~$ GeV$^3$ against a total magnitude of $p^2~m(p^2)$ $5.4~10^{-2}~$GeV$^3$ at $p \simeq 3~$GeV,$4.2~10^{-2}~$GeV$^3$ at $p \simeq 5.7~$GeV) but of course it would be better to have smaller errors to strengthen this conclusion \footnote{In fact,
the initial errors on the pseudoscalar vertex ($Z_P^{-1}$) are very small at large $p$, but the extraction of the residue increases them much.}.

\vskip 0.5 cm

\section{Conclusions}

\underline{General approach to the quark mass function} The method of circumventing the Wilson term artefact which affects the clover propagator quark mass function, by passing through the pseudoscalar vertex,  seems efficient, and yields a result no too far from the continuum.

\noindent \underline {OPE}. The present study give useful indications for the extraction of OPE power corrections, at least for elementary Green functions. In the regime of spontaneous chiral symmetry breaking, the quark mass function is in principle an exceptionally favorable case  for the study of the OPE power corrections, since 

1) the $1/p^2$ quark condensate contribution is the leading order in the chiral limit, 

2) the value of the condensate is well known by other methods. 

In fact, it is the best place for a lattice, nonperturbative, measurement of a Wilson coefficient. In practice, the present study strengthens the previous conclusion that an accurate OPE analysis is a very difficult task. 

The present case illustrates one of the difficulties, which will also appear in other cases according to the calculations of Chetyrkin and Maier~: the perturbative series giving the Wilson coefficient of the power term converges  at best very slowly~; it is difficult to work at sufficiently large momenta to have it converge better. In the present work, at the largest momenta, even with the huge improvement provided by the recent work of Chetyrkin and Maier (three loops), a discrepancy of a factor 2 seems to remain with the actual value of the condensate~; we find a fictitious $-<\bar\psi \psi>(2~{\mathrm GeV}) \simeq (4 \pm 0.6)~10^{-2}$ GeV$^3$. 
Equivalently, one can say that, practically, the OPE even with the very high orders calculated can work only at very high $p$. Effects of similar magnitudes are found by Chetyrkin and Maier in several other cases, concerning the $A^2$
vacuum condensate contribution to propagators. The situation about the OPE of composite operators is not known. 

This causes a difficulty for the program of renormalisation through the lattice measurement of MOM renormalisation constants.  This is especially true if we require a high precision of order of the ``percent", as often considered. Indeed, let us recall that the final aim of the method is to extract from the measured renormalisation constants {\bf their purely perturbative} part. Of course, in principle, we could work at a very large $p$ to kill completely the power corrections, and at a sufficiently large cutoff, in order that this large $p$ should not be affected by appreciable UV artefacts. However, from our findings, power corrections due to $<A^2>$ can be guaranteed to be $1 \%$ only around $p \simeq 10$ GeV. Then the task would be obviously very difficult.

The alternative method is to perform OPE fits at available momenta. Such fits must include first the known purely perturbative part, multiplied by the unknown $Z$, second the power corrections, which are predicted to be important at usual momenta, but also, as a third contribution, discretisation terms invariant under $O(4)$, which should be important as well at this degree of accuracy, and whose magnitude is unknown. The method is then to exploit the different behaviour of the various types of terms, as function of both $p$ and $\beta$. This method of extraction of $Z$ is then also difficult, when the form of the discretisation artefacts (as we have shown for the vector part of the propagator) as well as the functional form of the Wilson coefficient, which depends on high orders, are not known accurately. Moreover, we have not an independent knowledge of the $A^2$ condensate value in contrast to the quark condensate. Let us stress that the difficulty of OPE is presently manifest only in elementary, non gauge invariant, Green functions, but these are precisely the ones used for MOM normalisation conditions.

\noindent \underline {``Restoration of chiral symmetry"}. On the other hand, the study has revealed the unexpected and remarkable feature, that there is a general discontinuity of chiral properties, as obtained by the standard chiral extrapolation methods~: when the physical volume
passes through some critical value, chiral symmetry seems to be restored abruptly although we do not expect a real phase transition. The discontinuity affects simultaneously the quark mass function, the pion mass and the chiral condensate. Since the critical volume coincides with the one found for the phase transition to symmetry restoration advocated since several years by Neuberger and Narayanan at $N_C \to \infty$, we are tempted to establish a connection with it~: it could be a remnant of this at $N_C=3$. We stress the fact that this property is revealed only through a polynomial extrapolation from non negligible quark masses of $30~MeV$ or more (bare masses).

\noindent \underline {Alternative method for calculating the quark condensate}
As a by-product, our calculations suggest that the rather direct method of calculating the condensate through the chiral extrapolation of $m \int d^4 x <P_5(x) P_5(0)>$ is efficient. For our larger lattice, the results is very encouraging, considering the simplicity of the method~: $<\bar{\psi} \psi>_{\overline{MS}}(2~{\mathrm GeV}) =-(2.2 \pm 0.4)~ 10^{-2}~{\mathrm GeV}^3$.

\section{Acknowledgements}

\hspace{\parindent}We thank very much K. Chetyrkin and A. Maier to have communicated us the results of their calculation before publication, and for useful comments. We thank also very much Damir Becirevic for communicating us their plot at $\beta=6.0, 16^3,52$, and for many useful discussions on the subject since a very long time~; and Beno\^it Blossier for many useful suggestions and constant discussions. We thank C. Urbach for having informed us of their own work.

\appendix

\section{High orders of the Wilson coefficient of the condensate} \label{A}

\hspace{\parindent}From the calculations of Chetyrkin and Maier ( \cite{chetyrkin09110594} \cite{maier} ), we extract the following result at $N_f=0$ for Euclidean $p$~:
\begin{eqnarray}   \label{ordre4}
m(p^2)&=&16 \pi^2 /p^2~[\alpha_s /\pi (-\frac {1}{12})+ \nonumber \\
&&(\alpha_s /\pi)^2(-\frac {33}{64}) + \nonumber \\ 
&&(\alpha_s /\pi)^3 (-\frac {13 745}{3 072}+\frac {79}{1 536} \zeta(3))+\nonumber \\
&&(\alpha_s /\pi)^4(-\frac {26 331 733}{497 664}+\frac {2 236 285}{995 328}  \zeta(3)\nonumber \\
&&-\frac {79} {3072} \zeta(4)+\frac {12 166 325} {3 981 312} \zeta(5))](1+\Sigma(p)) <\bar \psi \psi>(p) 
\end{eqnarray}
$\alpha_s$ being taken at $p$. It is easily seen that the first order is as given by Politzer, and the second one as given by Pascual and de Rafael. We have obtained these numbers by the following manipulations. The authors consider the renormalised OPE of the quark propagator renormalised in the ${\overline{MS}}$ scheme, and write the contribution of the quark condensate as $C(q)_{\bar \psi \psi} <\bar\psi \psi>$. They find for the scalar part and vector part of the renormalised propagator $S=A(p)+B(p) \pslash$~: 
\begin{eqnarray}
A(p)=-m/p^2+C_{\bar \psi\psi}(p)<\bar\psi \psi> +...\nonumber \\
B(p)=-1/p^2+...
\end{eqnarray}
where dots denote $n>0$ orders in $\alpha$ or power corrections with power larger than $1/p^2$, or terms suppressed by $n>0$ powers of $m$. In $A(p)$, the tree order term $m$ is kept in a first step to indicate the conventions of the authors. We note that $m(p^2)$ as we calculate from the ratio of bare quantities is also the ratio of the renormalised quantities in any scheme, since the factor $Z_2$ applies to both terms of the fraction. Then, at $m=0$,~:
\begin{eqnarray}
m(p^2)=A(p)/B(p)=-p^2 C_{\bar \psi \psi}(p)<\bar\psi \psi> (1+\Sigma(p))
\end{eqnarray}
where $\Sigma(p)$ is the $\overline{MS}$ quark self-energy. Now, we pass to the Euclidean space by setting $p^2=-p_E^2$, or making the substitution $p^2 \to -p^2$. Then $C$ being read as $1/p^4 C'$, where $C'$ represents the radiative corrections and is a polynomial 
in $\alpha_s$, one ends on~:
\begin{eqnarray}
m(p^2)=1/p^2~C'(\alpha_s) <\bar\psi \psi> (1+\Sigma(p)) 
\end{eqnarray}
$C'$ and $<\bar\psi \psi>$ being negative, $m(p^2)>0$ as observed on the lattice. Finally, we set $\mu=p$. Then $\Sigma(p)$ is found to give a very small correction. We borrow the expression from Chetyrkin and Retey~: \cite{chetyrkin0007088}~:
\begin{eqnarray}
1+\Sigma(p)=1+1/4^2 (C_A C_F(41/4-3~\zeta(3))+C_F^2~(-5/8)) (\alpha_s/\pi)^2\nonumber \\
+1/4^3(C_A^2 C_F(159257/648-3139/24~\zeta(3)-69/16~\zeta(4)+165/4~\zeta(5))+\nonumber \\
C_A C_F^2(-997/24+44~\zeta(3)+6~\zeta(4)-20~\zeta(5))+C_F^3(-73/12))(\alpha_s/\pi)^3 \end{eqnarray}
or numerically~:
\begin{eqnarray} \label{sigma}
1+\Sigma(p)=1+1.59151 (\alpha_s/\pi)^2 + 23.2809 (\alpha_s/\pi)^3
\end{eqnarray}
This is $1.005$ for $\alpha_s=0.14$ ($p=5~{\mathrm GeV}$)

\end{document}